\documentclass[12pt]{article}

\usepackage{epsf,a4wide}

\begin{document}

\begin{flushright}
DESY-98-168 \\ CPHT-PC669-1098 \\ hep-ph/9811220
\end{flushright}

\begin{center}
\vskip 1.5\baselineskip
\textbf{\LARGE Exclusive electroproduction and \\[0.5ex]
  off-diagonal parton distributions\,\footnote{Talk given by M.D. at the
    Fourth Workshop on Quantum Chromodynamics, Paris, France, 1--6
    June 1998. To appear in the proceedings.}}
\vskip 2.5\baselineskip
M. Diehl \\[0.3\baselineskip]
\textit{DAPNIA/SPhN, CEA/Saclay, 91191 Gif sur Yvette, France \\
  present address: DESY, 22603 Hamburg, Germany} \\[\baselineskip]
T. Gousset \\[0.3\baselineskip]
\textit{SUBATECH, B.P. 20722, 44307 Nantes, France} \\[\baselineskip] 
B. Pire \\[0.3\baselineskip]
\textit{CPhT, Ecole Polytechnique, 91128 Palaiseau, France}
\vskip \baselineskip
\vskip 2\baselineskip
\textbf{Abstract} \\[0.5\baselineskip]
\parbox{0.9\textwidth}{Off-diagonal parton distributions occur in
  several hard exclusive reactions. They extend the study of hadron
  structure beyond what can be learned from ordinary distributions and
  have a particularly rich spin structure. The hard scattering
  subprocesses in electroproduction of mesons and of real photons
  satisfy helicity selection rules, which provide powerful tools to
  test leading-twist dominance at a given value of the hard scale.}
\vskip 1.5\baselineskip
\end{center}


\section{From diagonal to off-diagonal parton distributions}

\begin{figure}[t]
\begin{center}
  \leavevmode
  \epsfxsize 0.95\textwidth
  \epsfbox{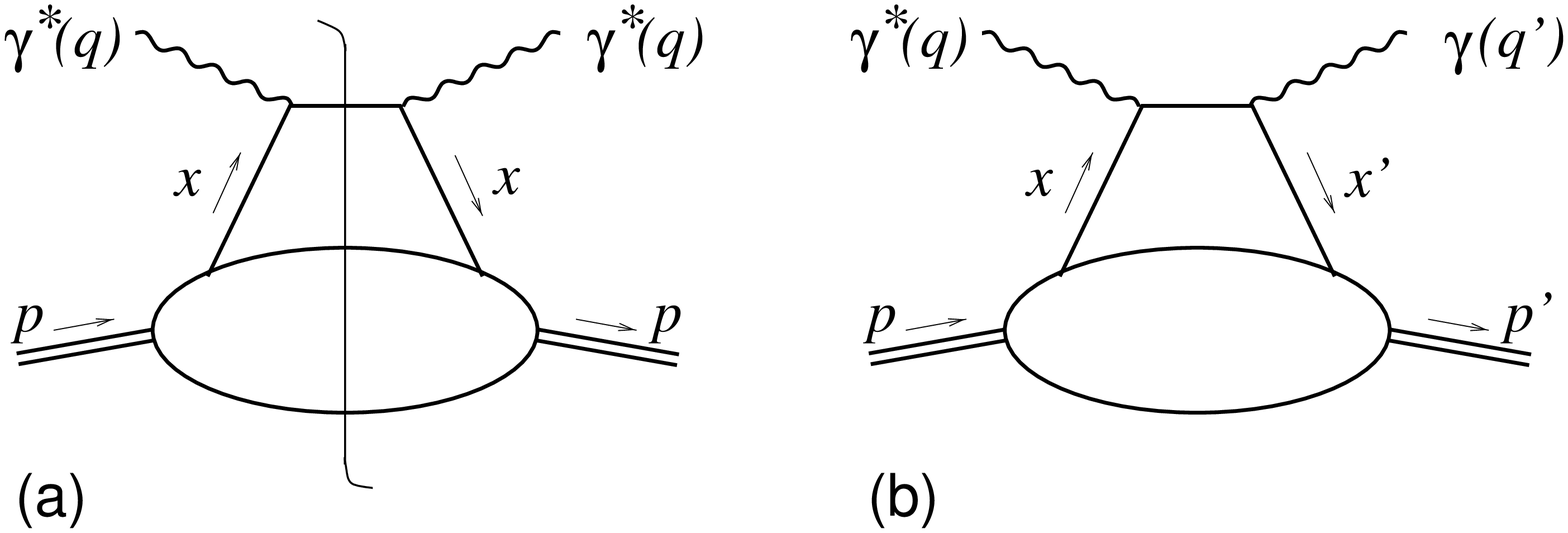}
\end{center}
\caption{\label{fig:compton} (a) Born level diagram for the forward
  Compton amplitude. The blob at the bottom denotes the quark or
  antiquark distribution in the proton and the vertical line the cut
  that gives the imaginary part of the amplitude. (b) Born level
  diagram for deep virtual Compton scattering. The blob now is
  described by an off-diagonal parton distribution.}  
\end{figure}

Off-diagonal (also called off-forward or nonforward) parton
distributions have enjoyed considerable interest since it was realised
that they allow the description of certain exclusive reactions in the
framework of QCD factorisation.\cite{OFPD,NFPD,CFS} A~good physics
example to see how they come about and how they are connected with the
usual, diagonal parton distributions, is the virtual Compton
scattering amplitude. Let us therefore start with the forward
amplitude $\gamma^\ast p \to \gamma^\ast p$, whose imaginary part
gives the cross section of inclusive deep inelastic scattering
$\gamma^\ast p \to X$ via the optical theorem. In the Bjorken limit
this can be calculated as a perturbative parton-photon scattering
times a parton distribution in the proton; the Born level diagram is
shown in Fig.~\ref{fig:compton} (a). The on-shell condition for the
parton line across the cut fixes the momentum fraction $x$ of the
parton in the proton to be $x_B$.

If we omit the cut of the diagram and replace the $\gamma^\ast$ on the
right-hand side with a real photon then we obtain
Fig.~\ref{fig:compton} (b), which describes the amplitude of the
exclusive process $\gamma^\ast p \to \gamma p$. Since there is now a
transfer of momentum between the proton on the left and on the right
the momentum fractions of the two partons attached to the lower blob
are no longer the same. This blob is described by an
\emph{off-diagonal} parton distribution, given as a Fourier
transformed matrix element of an operator between two \emph{different}
proton states. The factorisation shown in Fig.~\ref{fig:compton} (b)
holds in the regime of deep virtual Compton scattering (DVCS), where
in addition to the usual Bjorken limit of large $Q^2 = -q^2$ and fixed
$x_B = Q^2 /(2 p\cdot q)$ we require $t =(p-p')^2$ to be small and
fixed.\cite{fact} Even at Born level we now have to perform a loop
integral over $x$; external kinematics fixes the difference $x - x'
\approx x_B$, where both fractions $x$ and $x'$ refer to the incident
proton momentum $p$.

\begin{figure}[t]
\begin{center}
  \leavevmode
  \epsfxsize 0.5\textwidth
  \epsfbox{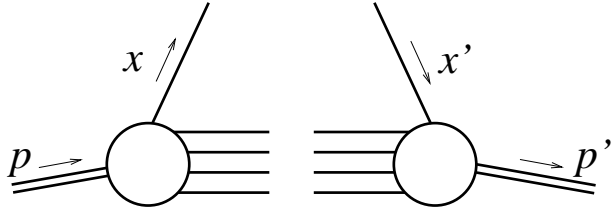}
\end{center}
\caption{\label{fig:interfere} Representation of an off-diagonal
  distribution in the range $x, x' >0$ as the interference between two
  different amplitudes for emitting a parton with a specified momentum
  fraction plus any number of spectators.}
\end{figure}

It can be shown that the proton-parton blob in Fig.~\ref{fig:compton}
(b) is effectively a \emph{cut} amplitude, although it appears in the
$\gamma^\ast p \to \gamma p$ amplitude which is not cut.\cite{order}
The off-diagonal distributions thus become equal to the usual,
diagonal ones in the limit where $p = p'$. Having a cut amplitude is
also important for the parton interpretation of these quantities. The
usual distributions represent the squared amplitude and thus the
probability for the proton emitting a parton with momentum fraction
$x$ and any number of spectators, summed over all spectator
configurations. In contrast the new distributions involve the
amplitude for parton emission with a fraction $x$ times the conjugated
amplitude with different fraction as shown in
Fig.~\ref{fig:interfere}; they therefore correspond to the
\emph{interference} between two different parton emission processes.

Such an ``interference experiment'' can be expected to provide new
insight into the proton structure. To see why let us take a numerical
example with $x_B = 0.3$ and $x = 0.4$. The parton emission at the
l.h.s.\ of Fig.~\ref{fig:interfere} takes then place at a value of $x$
where the usual valence quark distributions dominate over sea quarks
and spectator systems with few partons are preferred: having a large
number of partons in a proton state is difficult when most of the
proton momentum is already taken away by one parton. At the r.h.s.\ of
the figure the situation is just the opposite since the fraction $x' =
0.1$ is rather small, and a rather large number of spectators could be
accommodated. As one forces the spectator systems to be the same on
the left and on the right the proton is put into a quite unusual
dynamical situation, and its response will be sensitive to how the
proton wave function is made up from states with different parton
content and configuration.

There is another regime which does not occur in usual parton
distributions because it is due to the kinematic asymmetry of the two
protons: this is when $x > 0$ but $x' < 0$. Re-interpreting the parton
with negative momentum fraction $x'$ as an antiparton with positive
fraction $-x'$ one then has the picture of the proton $p$ emitting a
parton-antiparton pair and being left as proton $p'$. This is
reminiscent of a meson distribution amplitude, which describes the
transition between a meson and a parton-antiparton pair.

A special role is played by the point where $x = x_B$ and accordingly
$x' = 0$. Note that this configuration gives the imaginary part of the
DVCS amplitude in Fig.~\ref{fig:compton} (b) since putting the parton
line between the two photons on mass shell leads to the same
kinematical condition as in deep inelastic scattering,
Fig.~\ref{fig:compton} (a). For the parton distribution our numerical
example above has now been pushed to its extreme: while the parton on
the l.h.s.\ of the proton-parton blob has some finite fraction $x_B$
the one on the r.h.s\ is \emph{very} slow ($x' = 0$ should be
understood as up to small corrections because the condition $x = x_B$
is obtained by neglecting for instance the transverse momentum of the
parton in the proton).

\begin{figure}
\begin{center}
  \leavevmode
  \epsfxsize 0.9\textwidth
  \epsfbox{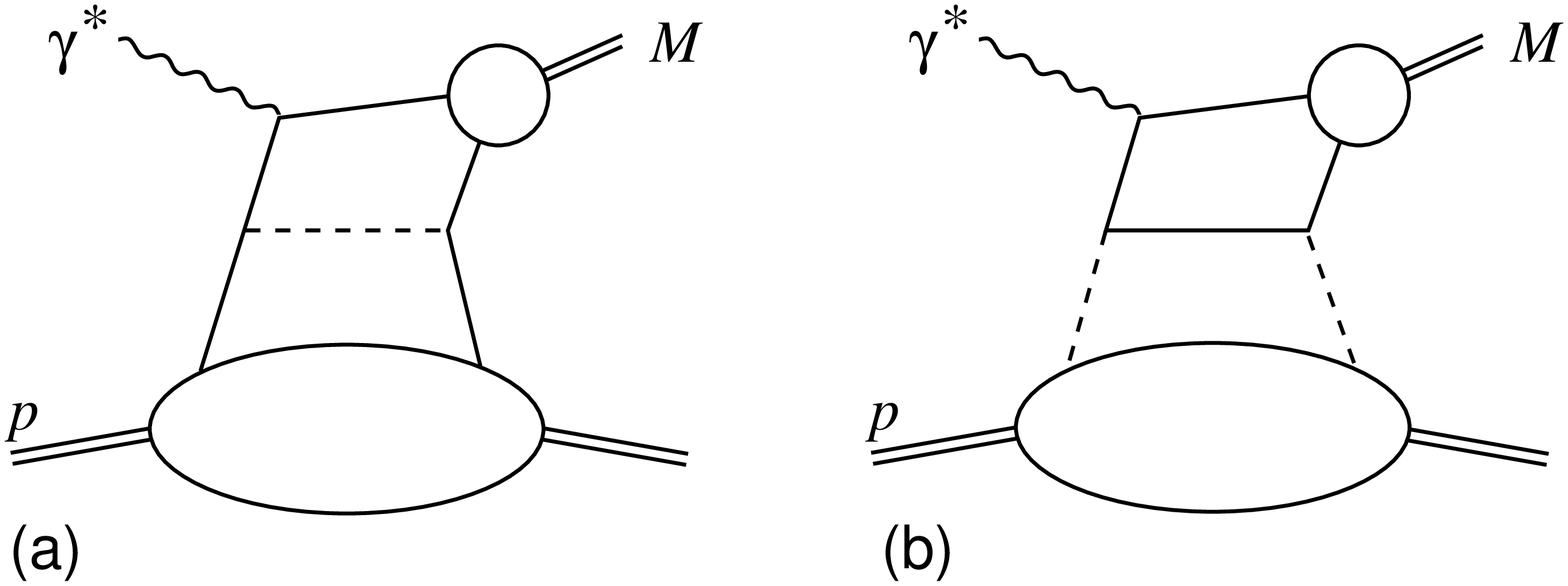}
\end{center}
\caption{\label{fig:meson} (a) One of the  diagrams for the production
  of a meson $M=\rho$, $\phi$, $\pi$, $\eta$, \ldots\ from a
  longitudinally polarised photon, involving an off-diagonal quark
  distribution. The dashed line denotes a gluon.  (b) The same with an
  off-diagonal gluon distribution. Here pseudoscalar mesons cannot be
  produced because of their quantum numbers.}
\end{figure}

Like their diagonal counterparts the off-diagonal distributions are
universal quantities, which do not only occur in the Compton
amplitude. It has been shown that the production of a meson from a
longitudinally polarised virtual photon can be described in the
limit of large $Q^2$ and small $t$ by a hard scattering, the
quark-antiquark distribution amplitude of the meson, and off-diagonal
quark or gluon distributions in the nucleon.\cite{CFS} This
factorisation is shown in Fig.~\ref{fig:meson}. Beyond Born level in
the hard scattering the off-diagonal gluon distribution also
contributes to DVCS, with the two gluons coupling to the two photons
via a quark loop.

As in other cases of QCD factorisation the mechanisms shown in
Fig.~\ref{fig:compton} and \ref{fig:meson} are dominant if the hard
scale $Q^2$ is sufficiently large, with corrections to them being
suppressed in powers of $1/Q$. A source for such corrections in DVCS
is for instance the hadronic component of the outgoing real photon,
which is not described by a pointlike coupling as in
Fig.~\ref{fig:compton} (b). One can expect this correction to be more
important in DVCS than in inclusive DIS, both photons in
Fig.~\ref{fig:compton} (a) being highly virtual. The values of $Q^2$
where power corrections can be neglected may thus well be different
for different processes, and the importance of such corrections will
have to be studied empirically for each case. On one hand one must
control such corrections if one wants to extract the off-diagonal
parton distributions that appear in the leading mechanism, and on the
other hand the relative importance of power corrections in similar
situations may guide us towards a better understanding of these
corrections themselves.

The most immediate test is of course that of scaling: up to
logarithmic terms the DVCS amplitude should become independent of
$Q^2$ at fixed $x_B$ and $t$, which exactly corresponds to Bjorken
scaling in inclusive DIS.  The production amplitude of a meson from a
longitudinal photon is predicted to behave as $1/Q$ times logarithms
in the Bjorken limit (for transverse photons, where the factorisation
of Fig.~\ref{fig:meson} does not hold, a decrease like $1/Q^2$ is
obtained from power counting arguments).\cite{CFS}

\section{Spin selection rules in the hard scattering}

We shall now see that more detailed tests for the dominance of the
leading twist contributions can be constructed from helicity selection
rules obeyed by the hard scattering.

In DVCS one finds that to leading order in $1/Q$ the helicities of the
initial and final state photons must be the same. In the hard
quark-photon or antiquark-photon scattering of Fig.~\ref{fig:compton}
(b) one can neglect the quark mass and the transverse momentum
transfer to the photon, both of which are small compared with $Q$.
Zero quark mass implies that the scattering cannot change the parton
helicity due to chiral invariance, and zero transverse momentum
transfer means collinear scattering, so that the angular momentum
along the collision axis is given by the particle helicities. Angular
momentum conservation then imposes that the photon cannot change its
helicity because the parton does not.

DVCS is measured in electroproduction, $e p \to e p \gamma$, with the
lepton $e$ being scattered at large angle and emitting a highly
virtual photon. The dependence of the $ep$ cross section on the
azimuthal angle $\varphi$ between the lepton plane and the
proton-photon plane, defined in the c.m.\ of the scattered proton and
photon, contains information on the virtual photon polarisation:
different helicity amplitudes of the $\gamma^\ast p \to \gamma p$
process give characteristic terms in the $\varphi$-distribution, which
can thus be used to test in a model-independent way our above helicity
selection rule.\cite{DGPR}

The phenomenology of $e p \to e p \gamma$ is very rich because Compton
scattering interferes with the Bethe-Heitler process, where the final
state photon is emitted from the lepton. Within the kinematical region
defining DVCS either of these processes can dominate, depending mainly
on the $ep$ collision energy. The interference term between the two
mechanisms is of particular interest: since the Bethe-Heitler process
is completely calculable it allows to investigate DVCS at
\emph{amplitude} level. Let us also mention that the imaginary part of
the DVCS amplitude, which as discussed above is very interesting in
connection with the dynamics of off-diagonal distributions, can be
accessed via the single spin asymmetry for longitudinal lepton beam
polarisation.

Along the same lines as for DVCS one can establish that in vector
meson production at leading order in $1/Q$ the meson carries
longitudinal polarisation.\cite{DGP} Together with the dominance of
longitudinal photons already mentioned this leaves us with a single
photon-meson helicity combination in the large-$Q^2$ limit. As in DVCS
one can investigate the virtual photon polarisation using the azimuth
between the leptonic and the hadronic plane, while the meson
polarisation is analysed by its decay distribution and offers a
further handle to see whether a given $Q^2$ is ``large''.

\section{Spin in the off-diagonal distributions}

Unlike a hard scattering subprocess the off-diagonal distributions,
which describe soft physics, are sensitive to a transverse deflection
of the proton by a few 100 MeV. This invalidates spin selection rules
whose derivation assumes a collinear situation. The result is a richer
structure in proton and parton spin than the one for diagonal
distributions.\cite{Hood}

\subsection{Quark and antiquark distributions}

If neither parton nor proton helicity are flipped there are two
independent proton-parton amplitudes, with the parton and proton spins
being parallel or antiparallel. Taking their sum (difference) one
obtains parton spin independent (dependent) distributions, which in
the diagonal limit become the usual distributions $q(x)$ and $\Delta
q(x)$, resp. Two more off-diagonal distributions correspond to the
case where the proton spin is changed. For them the sum of proton and
parton spins is not conserved and angular momentum conservation is
restored by orbital angular momentum. That these distributions carry
indeed information on orbital angular momentum is also seen at the
level of the sum rules they obey.\cite{OFPD}

In DVCS all four of these distributions contribute to the cross
section, even when the polarisations of incident and scattered proton
are not measured. Knowledge of proton polarisation \emph{is} however
required if one wants to disentangle the different distributions from
the cross section. For meson production the situation is somewhat
different: the production of vector mesons only picks out the parton
spin independent distributions and pseudoscalar meson production only
selects the parton spin dependent ones.

There are also chiral-odd distributions, where the helicity of the
quark or antiquark is flipped; one of them becomes equal to the
so-called transversity distribution in the diagonal limit. The
original hope~\cite{CFS} that they may be measurable in the production
of a transverse vector meson from a longitudinal photon has
unfortunately not come true, because as mentioned above the hard
scattering process forbids such a transition.\cite{DGP,Mank}

\subsection{Gluon distributions}

For gluons there are again amplitudes corresponding to the usual spin
independent and spin dependent gluons distributions, as well as rather
exotic \emph{gluon helicity flip} amplitudes, the latter changing the
gluon spin by two units.

Vector meson production at large $Q^2$ only involves the gluon spin
independent distributions, at least to leading order in
$\alpha_s$.~\cite{Vaentt} In DVCS the gluon helicity flip
distributions do occur at the level of radiative
corrections.\cite{DGPR,Hood} For gluons there is no equivalent of
chirality conservation, which we used for quarks, and the hard
scattering compensates the gluon helicity flip by also flipping the
photon helicity by two units. Our above helicity selection rule for
DVCS thus acquires a distinct $O(\alpha_s)$ correction: gluon exchange
allows for transitions between two different transverse polarisations,
but not from a longitudinal to a transverse photon.  This has again a
clear signature in the distribution of the azimuth $\varphi$ discussed
above.

\section*{Acknowledgements} We gratefully acknowledge discussions
with J.P. Ralston and O.V. Teryaev. This work has been partially
funded through the European TMR Contract No.~FMRX-CT96-0008: Hadronic
Physics with High Energy Electromagnetic Probes. SUBATECH is Unit\'e
mixte 6457 de l'Universit\'e de Nantes, de l'Ecole des Mines de Nantes
et de l'IN2P3/CNRS, and CPhT is Unit\'e mixte 7644 du CNRS.


\begin{thebibliography}{99}
  
\bibitem{OFPD} X. Ji, Phys.\ Rev.\ Lett.\ {\bf 78}, 610 (1997); Phys.\ 
  Rev.\ {\bf D55}, 7114 (1997).

\bibitem{NFPD} A.V. Radyushkin, Phys.\ Rev.\ {\bf D56}, 5524 (1997).
  
\bibitem{CFS} J.C. Collins, L. Frankfurt and M. Strikman, Phys.\ Rev.\
{\bf D56}, 2982 (1997).

\bibitem{fact} X. Ji and J. Osborne, Phys.\ Rev.\ {\bf D58}, 094018
  (1998); \\ J.C. Collins and A. Freund, hep-ph/9801262.
  
\bibitem{order} M. Diehl and T. Gousset, Phys.\ Lett.\ {\bf B428}, 359
  (1998); \\ R.L. Jaffe, Nucl.\ Phys.\ {\bf B229}, 205 (1983).

\bibitem{DGPR} M. Diehl, T. Gousset, B. Pire and J.P. Ralston, Phys.\
Lett.\ {\bf B411}, 193 (1997).

\bibitem{DGP} M. Diehl, T. Gousset and B. Pire, hep-ph/9808479.

\bibitem{Hood} P. Hoodbhoy and X. Ji, Phys.\ Rev.\ {\bf D58}, 054006
(1998).

\bibitem{Mank} L. Mankiewicz, G. Piller and T. Weigl, Eur.\ Phys.\ J.\
{\bf C5}, 119 (1998).

\bibitem{Vaentt} M. V\"anttinen and  L. Mankiewicz, Phys.\ Lett.\ {\bf
    B434}, 141 (1998).

\end{thebibliography}
\end{document}